\documentclass[journal=amlccd,manuscript=letter,
layout=twocolumn]{achemso}

\usepackage{amsmath}    % need for subequations
\usepackage{color}      % use if color is used in text
\usepackage{graphicx}   % need for figures
\usepackage{bm}
\usepackage{achemso}

\title{Threading Dynamics of Ring Polymers in a Gel}

\author{Davide Michieletto}
\affiliation{Department of Physics and Complexity Centre, University of Warwick, Coventry CV4 7AL, United Kingdom}
\author{Davide Marenduzzo}
\affiliation{School of Physics and Astronomy, University of Edinburgh, Mayfield Road, Edinburgh EH9 3JZ, Scotland, United Kingdom}
\author{Enzo Orlandini}
\affiliation{Dipartimento di Fisica e Astronomia and Sezione INFN, Universit\`a di Padova, Via Marzolo 8, 35131 Padova, Italy}
\author{Gareth P. Alexander}
\author{Matthew S. Turner} 
\email{m.s.turner@warwick.ac.uk}
\affiliation{Department of Physics and Complexity Centre, University of Warwick, Coventry CV4 7AL, United Kingdom}
\date{}

\AbstractOn

\begin{document}

%%%%%%%%%%%%%%%%%%%%%%%%%%%%%%%%%%%%%%%%%%%%%%%%%%%%%%%%%%%%%%%%%%%%%%%%%%%%%%
%\begin{tocentry}
%\begin{wrapfigure}{l}{0.15 \textwidth}
%\vspace{-0.1 cm}
%\hspace*{-0.15 cm}
%\includegraphics[scale=0.052]{Fig1_tot_seqBis.pdf}
%\vspace{-0.8 cm}
%\end{wrapfigure}
%\end{tocentry}
%%%%%%%%%%%%%%%%%%%%%%%%%%%%%%%%%%%%%%%%%%%%%%%%%%%%%%%%%%%%%%%%%%%%%%%%%%%%%%

\begin{abstract}
We perform large scale three-dimensional molecular dynamics simulations of unlinked and unknotted ring polymers diffusing through a background gel, here a three-dimensional cubic lattice. Taking advantage of this architecture, we propose a new method to unambiguously identify and quantify inter-ring threadings (penetrations) and to relate these to the dynamics of the ring polymers.  We find that both the number and the persistence time of the threadings increase with the length of the chains, ultimately leading to a percolating network of inter-ring penetrations. We discuss the implications of these findings for the possible emergence of a topological jammed state  of very long rings.
\end{abstract}

\begin{figure*}
\includegraphics[scale=0.09]{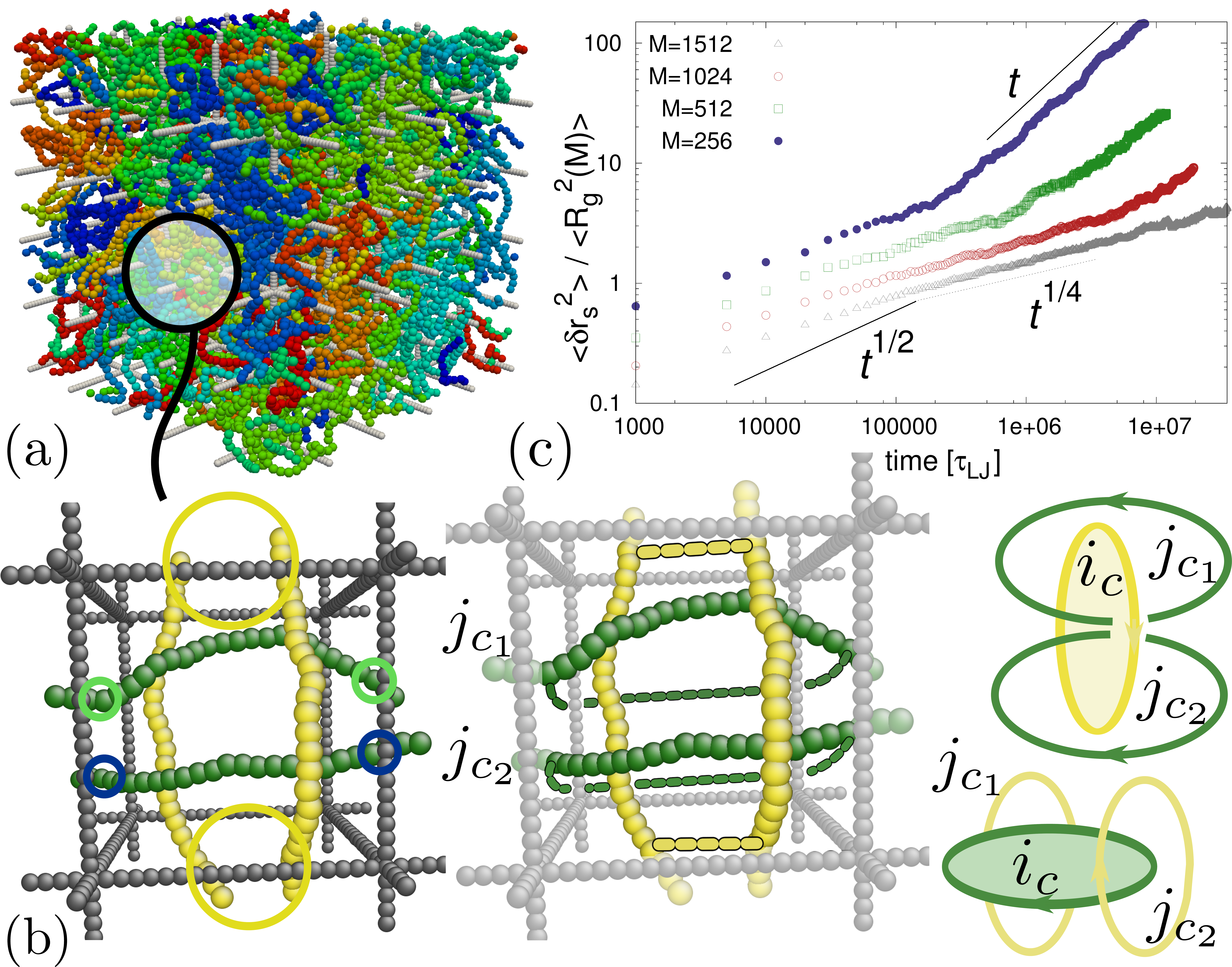}
\caption{\small (color online). (a) Snapshot of a system with $N=50$ chains of length $M=256$. The gel lattice (grey) can be seen to be interpenetrated by ring polymers (various colours). We use periodic boundary conditions so there are no real ends to either the gel or the polymers. (b) Sketch showing our procedure for identifying rings that thread in a given unit cell $c$ of the gel lattice. Here the green strands of chain $j$ passing through a face of the cell (circled) are separately closed to form two loops $j_{c_1}$ and $j_{c_2}$. Each are topologically linked with the yellow contour $i_c$, a unique contraction of chain $i$ formed by connecting the points that pass through the faces of the unit cell (circled). In contrast the green ring is not threaded by the yellow. See text for details. (c) Segmental mean squared displacement of the rings $\langle \delta r^2_s \rangle = \langle [\bm{r}_i(t)-\bm{r}_i(0)]^2 \rangle$ scaled by $\langle R^2_g(M)\rangle$ and plotted against time in units of the Lennard-Jones time $\tau_{LJ} =$ $\sigma$ $(m/\epsilon)^{1/2}$ where $m$ is the mass of the beads and the target temperature is $T=\epsilon/k_B$. The gel structure is here thinned for clarity.}
\label{fig:FIG}
\end{figure*}

Understanding the dynamical and rheological properties of solutions of long polymers is of primary importance in several area of soft matter, material science and biophysics~\cite{Everaers2004} .
The dynamics of linear polymers in the melt is now understood using the tube and reptation models~\cite{Doi1988,Gennes1979a} . These models take advantage of the topological constraint represented by the non-crossability of the chains to describe the diffusion of the polymers along their own primitive path, by relaxing the free ends. By contrast, the dynamics of ring polymers, which have no free ends, can differ markedly from those of their linear or branched cousins in the melt~\cite{Rubinstein1986,Grosberg1993,Obukhov1994,Deutsch1999,Muller2000,Kapnistos2008,Suzuki2008,Vettorel2009,Sakaue2011,Robertson2006,Robertson2007} , 
involving fundamentally different modes of stress relaxation~\cite{Kapnistos2008} , significantly different diffusion constants~\cite{Robertson2006,Robertson2007} and a crossover to free diffusion that occurs only once they travelled many times their own size $\langle R^2_g \rangle^{1/2}$~\cite{Halverson2011a} . Inter-ring penetrations, or ``threadings'', have previously been speculated to play some role in ring dynamics~\cite{Klein1986,Kapnistos2008,Halverson2011,Halverson2011a, Vettorel2009,Lo2013, Bernabei2013} , although no methodology to define or identify them yet exists. The goal of this work is to study a system in which we can quantify these threadings and their effect on the long-time dynamics of a concentrated solution of ring polymers. At present its not possible to identify such threadings in the melt.  Here we focus our attention on a system that is rather different from a melt of rings: We study a concentrated solution of ring polymers embedded in a physical gel which, for simplicity, we model as a rigid cubic lattice, see Fig.~\ref{fig:FIG}(a).
As we show below this system is well suited to the study of inter-ring penetrations (threadings). It is also highly accessible from an experimental point of view, as it resembles the classical setup used for gel electrophoresis~\cite{Deutsch1988,Alon1997} of plasmid DNA rings, except that the polymer concentration is taken above overlap and the gel is prepared in order to have a pore size comparable to the polymers' Kuhn length. 
In this Letter we introduce for the first time a quantitative measure of inter-ring threadings for polymers diffusing in a background gel by using a novel algorithm that employs the background gel as a reference frame. We can then study the consequences of this on their dynamics. We show that the number of threadings grows linearly with the degree of polymerisation $M$ of the chain and that this leads to the emergence of an extended directed network of threadings that includes of order all rings. This network of threadings is associated with the onset of very slow dynamics and we show that the un-threading process drives the emergence of a significant slowing down of the longest rings we are able to study. Finally, we speculate that such a threading-rich state may be a precursor of a {\it topological jammed} state for even longer chains, as these threadings provide long-lived ``pinning'' sites that represent severe topological constraints in the polymers' diffusion.
We study unlinked and unknotted ring polymers diffusing through a background gel (see Fig.~\ref{fig:FIG}(a) and~\ref{fig:FIG}(b)), formed by a perfect cubic lattice, \textit{i.e.} without dangling ends. The novel aspect of this work is in how we are able to identify the role of inter-ring threadings. By neglecting fluctuations of the gel we can direct computational resources most effectively towards simulating the dynamics of the rings themselves. We use a molecular dynamics engine (LAMMPS) to model the Langevin dynamics at fixed volume and constant temperature of $N$ polymer rings with length $M$ moving inside a three dimensional cubic lattice of total linear size $L$ and lattice spacing $l$.
The ring monomer density is kept constant for all the systems at $\rho=N M/L^3=0.1$ $\sigma^{-3}$ (on top of this, the density of the gel is $0.06$ $\sigma^{-3}$). The well established Kremer-Grest model~\cite{Kremer1990} is used to simulate worm-like chains with non-crossability constraint and excluded volume interaction (see S.I. for simulation details).  
 
The Kuhn length provides a natural choice for the lattice spacing of the gel, $l=l_K=10$ $\sigma$, where $\sigma$ is the nominal size of a bead composing the polymers. By making this choice we assume that the mesh size of our idealised gel corresponds to that of a moderately concentrated agarose gel~\cite{Fatin-Rouge2004} , or appropriate DNA hydro-gel architecture~\cite{Um2006,Park2009,Lee2012} . From a physical perspective, lattice spacings much greater than the Kuhn length can produce a gel so sparse that the rings rarely encounter it. This maps the problem back onto the melt, a different system from the one we study here, and one that is less well suited to the study of inter-ring penetrations. Alternatively, for lattice spacings much shorter than the Kuhn length threadings will ultimately be suppressed by steric effects. Also, the simulation will include an increasingly large fraction of passive gel monomers, which tend to increase the volume fraction of the system and hence limit the concentration of rings that can be studied efficiently using LAMMPS. The choice of $l=l_K=10$ $\sigma$ is a natural compromise and corresponds to systems that can have numerous threadings, as we show below. This is due, at least in part, to the fact that the polymers are forced to spread across many unit cells, and within each cell the polymers are stiff.

We study threadings as {\sl local} properties of the conformation of rings; the {\sl global} topology of the rings remains unlinked from both other rings and the gel. Here the gel architecture provides a natural local volume -- a single unit cell -- within which threading of one ring by another can be identified; no corresponding method exists for the melt. Each polymer enters and exits a given cell through its faces. The unique topological characteristic of ring polymers, unlinked from the cubic lattice, is that each time the contour passes out through a face of any given unit cell, labelled $c$, this must be accompanied by a returning passage back through the same cell face.  The threading of polymer $i$ by polymer $j$ within cell $c$ can then be defined as follows: First a contraction of ring $i$ is formed by sequentially connecting the points where it passes through any face of cell $c$ by straight lines, as illustrated by the dashed (yellow) lines in Fig.~\ref{fig:FIG}(b). This creates a closed loop (or link) $i_c$ contained entirely within cell $c$ and its bounding faces. In this way we use the gel to identify threadings as local configurations in which the conformation of the ring outside of the chosen cell is unimportant. Next we consider each of the strands, labelled by $j_c$, of a different polymer $j$ in the same cell $c$. These strands connect a single entry and exit point through the faces of $c$. We now close the ends of each strand outside the cell to form a closed loop. We then compute the linking number of the loop thereby created from each $j_c$ with $i_c$. This will be non-zero if, and only if, ring $i$ is threaded by that strand of ring $j$. For instance, the two strands of the green ring in Fig.~\ref{fig:FIG}(b) are threading the yellow ring, since the absolute value of the linking numbers of each of these (after closure) with the closed yellow loop are equal to one. 
We define the local threading of ring $i$ by ring $j$ in cell $c$ at time $t$ by $Th_c(i,j;t) = \tfrac{1}{2}\sum_{j_c}|Lk(i_c,j_c;t)|$ -- equal to $1$ for the example shown in Fig.~\ref{fig:FIG}(b) -- and the total threadings between these rings by summing this over all cells 
\begin{equation}
Th(i,j;t)  =  \sum_{c} \sum_{j_c} \dfrac{\left| Lk_c(i_c,j_c;t) \right|}{2} \label{eq:Th}
\end{equation}
This procedure is perfectly well defined, even when rings enter and leave through the same faces of the cell (see also S.I.). We emphasise that this is taken to be a {\it definition} of threading. It is necessarily a strictly local measure, on the scale of the cell volume. If the cell volume is increased no threadings will eventually be recorded since rings in the melt are unlinked by construction.

We assign a {\it passive} threading of ring $i$ by ring $j$ when $Th_c(i,j;t)=1$ and $Th_c(j,i;t)=0$ and an {\it active} threading when $Th_c(i,j;t)=0$ and $Th_c(j,i;t)=1$. For example, in Fig.~\ref{fig:FIG}(b), the yellow ring is passively threaded by the green one, which is actively threading the yellow one. 

\begin{figure}
\includegraphics[scale=0.36]{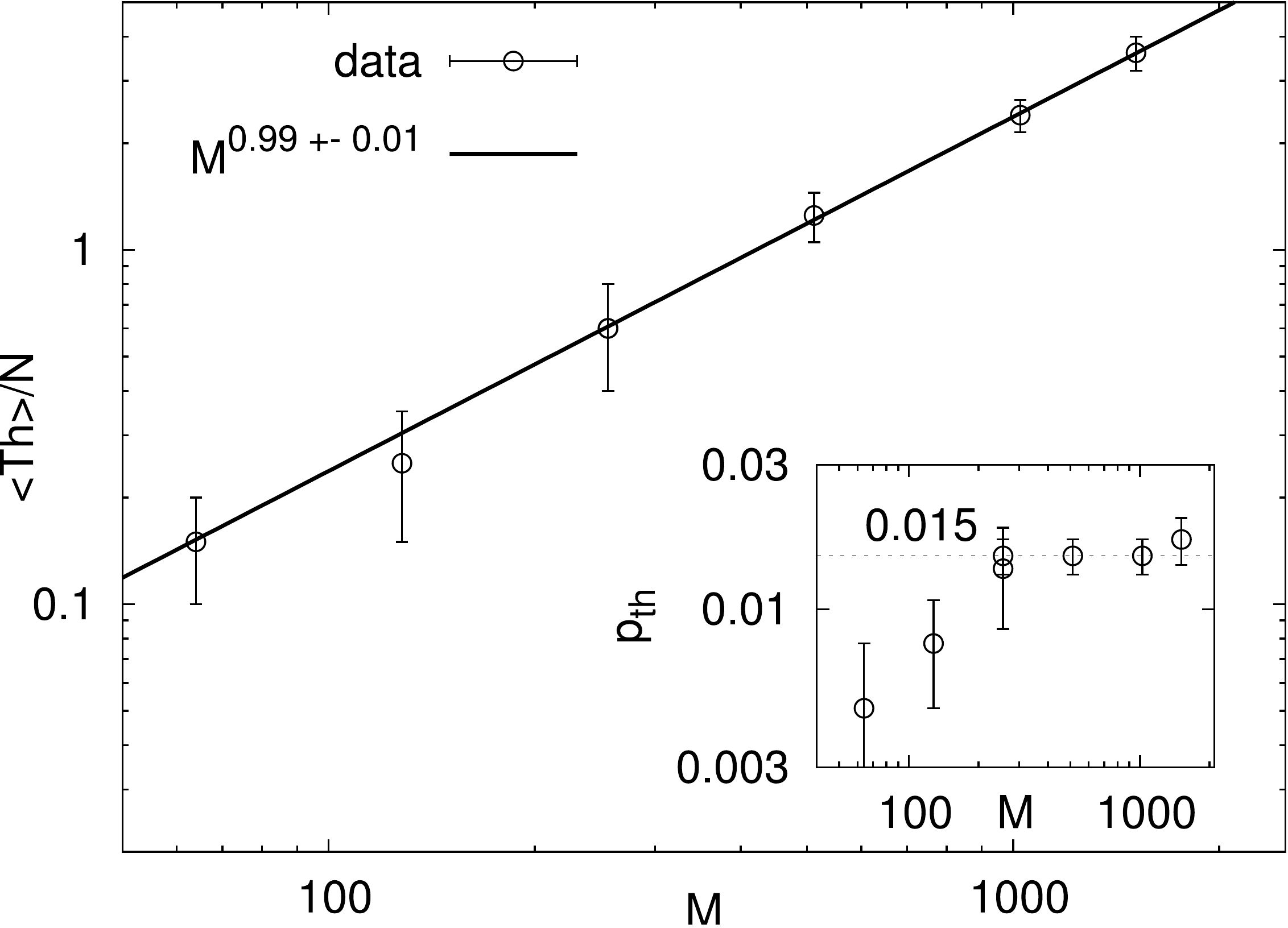}
\caption{\small Number of threadings $\langle Th \rangle$ per chain as a function of the length of the chains $M$. In the inset we plot $p_{th}$, crudely the probability of threading in a cell containing two different chains, as computed in eq.~(3). See text for details.} 
\label{fig:ThM}
\end{figure}

The equilibrium average $\langle Th(i,j;t) \rangle_{i,j,t}/N \equiv \langle Th \rangle/N$ is the number of threadings per chain and is found to scale extensively with the ring length $M$. This may be related to the fact that the number of  cells visited by each chain also grows linearly with $M$ (see Fig.~\ref{fig:ThM}).

We claim that the existence of these penetrations influences the dynamics of the rings by pinning chains' segments. A measure of this is given by their time-correlation function

\begin{equation}
P_p(t) =    \left\langle \dfrac{ \sum_j p(i,j; t_0)p(i,j; t_0+t)}{\sum_j p(i,j; t_0)} \right\rangle_{i,t_0}  \label{eq:PenPers}
\end{equation} 

where $p(i,j;t)=1$ if ring $j$ is penetrating (threading) ring $i$ at time $t$ and $0$ otherwise. 

\begin{figure}
\includegraphics[scale=0.31]{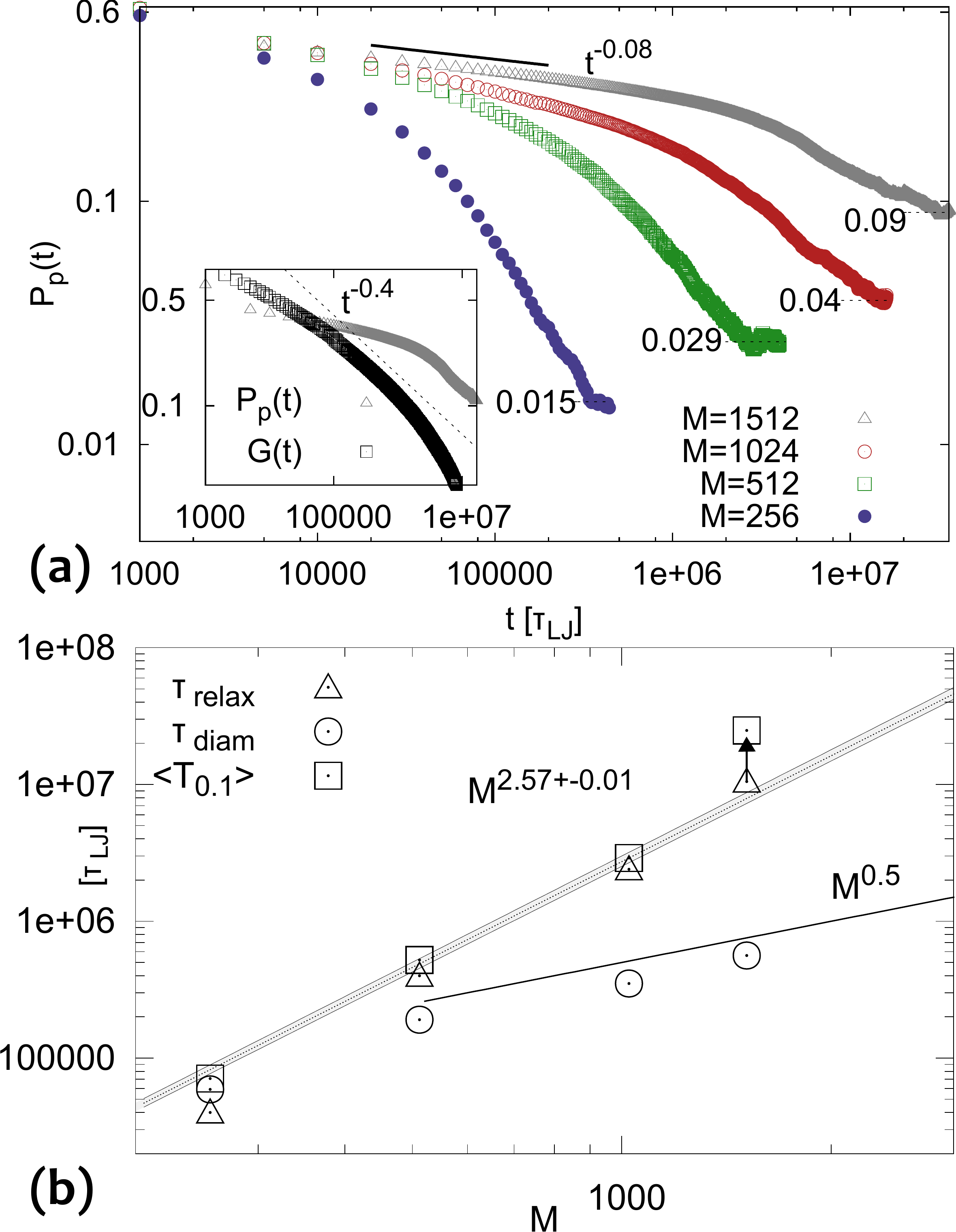}  
\caption{\small (color online). (a) Time-correlation function $P_p(t)$ of the penetrations as in eq.~(2). The inset compares the relaxation of the stress carried by the rings, through the modulus $G(t)$, with $P_p(t)$ for the system with longest rings showing that the spatial stress relaxes more quickly than the threadings. (b) Compares three dynamic relaxation times, defined as $\tau_{relax} \equiv \langle R^2_g \rangle/6D_{CM}$, $\tau_{diam} \equiv \int_0^\infty C_{diam}(t)dt$ ($C_{diam}(t)$ is the time auto-correlation function of the diameter vector $\vec{d}(t)$ that joins opposite beads along the rings' contour (see S.I. for details)) and $\langle T_{0.1}\rangle$, given by the solution of $P_p(T_{0.1}) \equiv 0.1$. The arrow on top of the data point showing $\tau_{relax}$ for the longest rings $M=1512$ indicates that this represents a lower bound: the crossover to diffusive motion has not yet occurred at the longest computationally accessible times. The unthreading times $\langle T_{0.1}\rangle$ for $M=256, 512, 1024$ follow the power law shown and are  close to the corresponding values of $\tau_{relax}$. The shaded region delimits the confidence bounds expected for the final data point for the unthreading time $\langle T_{0.1}\rangle$ at $M=1512$, were it to continue to follow this power law. This point is approximately 30 standard deviations outside the confidence interval, consistent with a dramatic slowing-down due to the development of a strongly connected network of inter-ring penetrations (threadings).}
\label{fig:PersPen}
\end{figure}

For large ring length $M$, $P_p(t)$ tends to flatten, resembling a plateau, before relaxing to a constant value $P_p(t \rightarrow \infty)=\langle p(i,j;t_0)\rangle_{i,j,t_0}$, this being the mean probability that two different, randomly chosen chains are penetrating. By using a mean-field argument the probability of threading between two chains in any cell that they both occupy $p_{th}$ can be approximated as the total probability that they thread divided by the number of shared cells that they both occupy $N_{sc}$
\begin{equation}
p_{th} = P_p(t \rightarrow \infty)/N_{sc}   \label{eq:pth}
\end{equation} 
We combine our measurement of $N_{sc}$ and $P_p(t \rightarrow \infty)$ in order to compute $p_{th}$, which is plotted in the inset of Fig.~\ref{fig:ThM}. 

We now compare the relaxation of threading with that of the modulus $G(t)$ for the stress carried by the rings, here computed as
\begin{equation}
G(t) =  \left\langle \dfrac{  \sum_c g(i,c; t_0)g(i,c; t_0+t)}{\sum_c  g(i,c;t_0)} \right\rangle_{i, t_0}  \label{eq:StrssRlx}
\end{equation} 
where $g(i,c;t)=1$ if ring $i$ visits cell $c$ at time $t$ and $0$ otherwise. While $G(t)$ is a standard quantity in polymer science the rheological response of the rings will be ``mixed" with that of the gel. This could lead to possible complications in isolating the rheological response of the rings alone. We therefore propose that diffusion of labelled tracer rings may be the most effective experimental probe of their dynamics.  From the inset of Fig.~\ref{fig:PersPen}(a) it is clear that the spatial stress relaxes more quickly than the threadings. This is consistent with the fact that one ring, penetrated by another in any particular cell, can independently relax the stress it carries in all other cells. As reported previously~\cite{Obukhov1994,Kapnistos2008,Halverson2011a,Pasquino2013} , the stress is found to relax faster than for melts of linear polymers, lacking any glassy plateau.  Long-lived penetrations may be responsible for the fact that the segmental mean square displacement of a ring only starts to freely diffuse ($\langle \delta r^2_s \rangle\sim t$) after the ring itself has moved many times $R_g$,  see Fig.~\ref{fig:FIG}(c). This is in contrast to linear polymers, where there are no penetrations and the same crossover is on the order of $R_g$ ~\cite{Halverson2011a} . We infer that free diffusion can only occur when the most persistent penetrations have relaxed, on the time-scales shown in Fig.~\ref{fig:PersPen}(a).

\begin{figure*}
\includegraphics[scale=0.29]{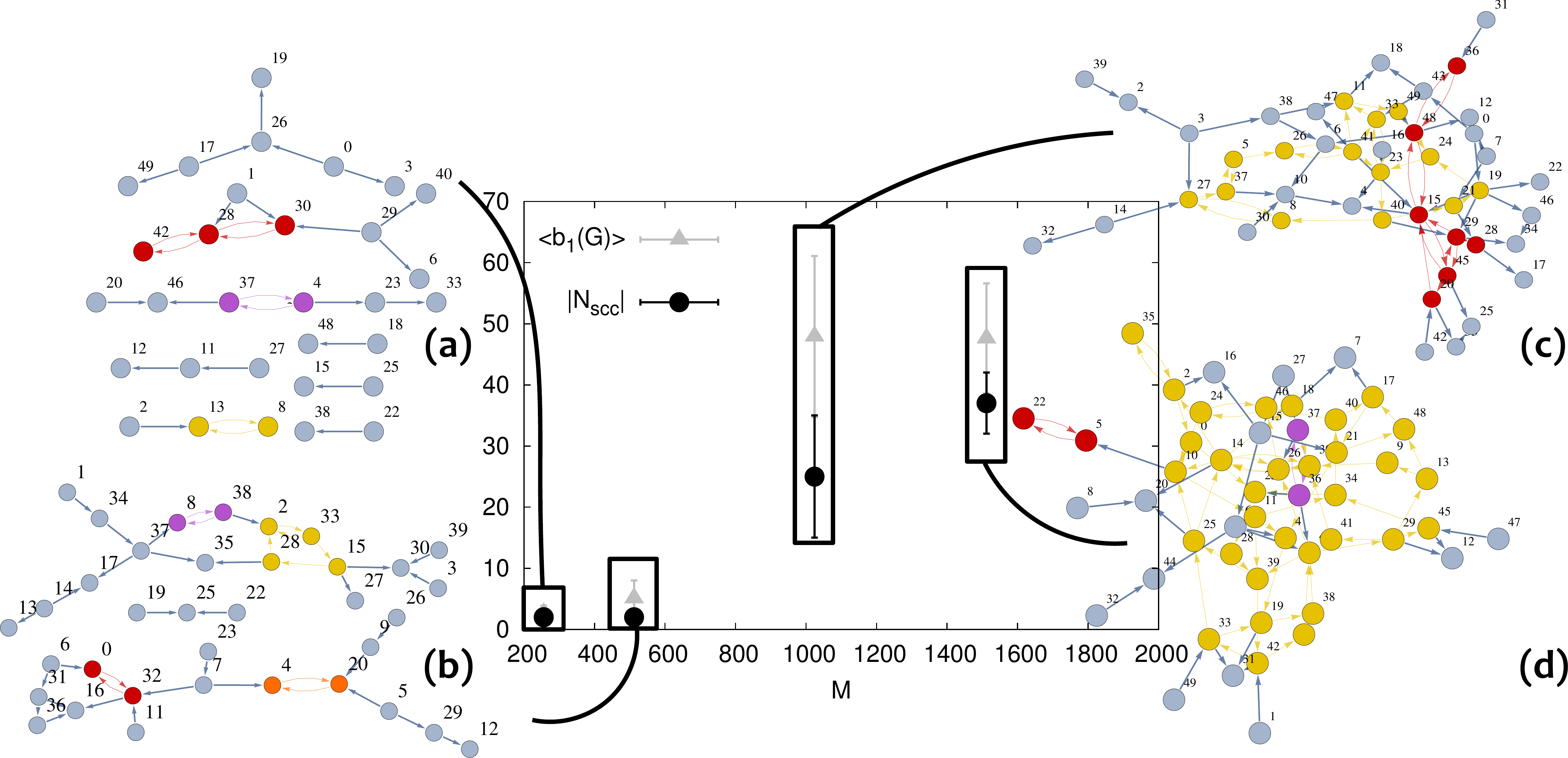}
\caption{\small (color online). Betti number $b_1(G)$ (see S.I.) and size of the largest strong connected component $|N_{scc}(G)|$ computed taking the time average after equilibration. Snapshots of the graphs $G({\cal E},{\cal V})$ corresponding to $M=256$ (a), $512$ (b), $1024$ (c) and $1512$ (d) beads. The colors highlight the strongly connected components in the graphs (see text for details).}
\label{fig:ConCompAllbis}
\end{figure*}
Fig.~\ref{fig:PersPen}(b) shows that chain re-orientation is fast. For $M=256$ it is comparable to the unthreading and diffusive relaxation times but it is much faster for $M \geq 512$. It may be significant that this is at the same point that significant clusters of interpenetrated rings start to appear, see Fig.~\ref{fig:ConCompAllbis}. These clusters do little to inhibit intra-chain re-organisation but the network of mutual pinning seems to be associated with the slowing of unthreading and diffusive relaxation. Fig.~\ref{fig:PersPen}(b) also shows that the diffusive relaxation time closely follows the un-threading time, supporting the hypothesis that free diffusion is possible only once the most persistent threadings are lost. In addition, we show that for the longest rings the penetrations relax much more slowly. We associate this with the emergence of strongly connected components in the network of inter-ring penetrations, see Fig.~\ref{fig:ConCompAllbis}.  This is consistent with figures~\ref{fig:ThM} and~\ref{fig:PersPen}(a) showing that, as we increase the length of the rings at fixed density, there is a corresponding increase in the number of threadings which also become more long-lived. In the percolating, long-lived cluster of inter-threading rings that emerges the motion of each ring is strongly constrained by its passive threadings. 

In order to quantify the network of penetrations we define a directed graph $G = G(\mathcal{E},\mathcal{V})$ where $\mathcal{V}$ is the set of vertices, a subset of the set of $N$ rings in the system, and $\mathcal{E}$ is the set of directed edges from ring $j$ to ring $i$, which represent the threadings of ring $j$ through ring $i$. We keep track of the time-evolution of the network via the matrix $Th(i,j;t)$ and quantify the emergence of extended structures of inter-threading rings by using the size of the largest strongly connected component $|N_{scc}|$ and the first Betti number $b_1(G)$ (see S.I.). 
As one can notice from Fig.~\ref{fig:ConCompAllbis}, most connected structures for short rings are formed by only two  mutually threading rings while, for longer rings, these eventually contain $\mathcal{O}(N)$ vertices, signifying the emergence of a percolating cluster of inter-penetrating rings which scales with the size of the system. We claim that such static transition in the structure is related to a correspondent dynamic transition, whose effect can be observed in the significant deviation showed in Fig.~\ref{fig:PersPen}(b).

In summary, we have employed a new method to quantify inter-ring threadings and to relate these to their dynamics in a background gel. Our findings suggest that inter-ring penetrations become more important as the length of the rings increases and that the dynamics of the polymers slows correspondingly. We highlight the existence of strongly connected components in the network of inter-penetrating rings and show that such components grow as the length of the rings is increased with a cluster of size $\mathcal{O}(N)$ inter-penetrating rings emerging for the longest chains we study. Together with the result that $Th(M)/N \sim M$ we speculate that  the dynamics is likely to be strongly suppressed for even longer rings. The term {\em topologically jammed} state might be used to describe this highly inter-penetrated state of matter which could have the unusual property that the dynamics could appear glassy, or at least slowed, well above the classical glass temperature $T_g$ for the polymer itself and hence without appreciable loss of microscopic mobility. As the molecular weight of the ring polymers increases, the dynamics is expected to dramatically slow down due to the topology of the polymers, which must unthread one-another in a particular order. This is rather different to the corresponding dynamics of linear polymers. This state would also inherit the well-known universality class of polymer physics and it would therefore seem to offer a novel framework in which to study a jamming transition from a topological perspective.

\begin{acknowledgement}
DMi acknowledges the support from the Complexity Science Doctoral Training Centre at the University of Warwick with funding provided by the EPSRC (EP/E501311). EO acknowledge financial support from the Italian ministry of education grant PRIN 2010HXAW77. We also acknowledge the support of EPSRC to DMa, EP/I034661/1, and MST, EP/1005439/1, the latter funding a Leadership Fellowship. The computing facilities were provided by the Centre for Scientific Computing of the University of Warwick with support from the Science Research Investment Fund
\end{acknowledgement}

\textbf{Supporting Information Available}  This information is available free of charge via the Internet at http://pubs.acs.org/journal/amlccd.

%\bibliography{ThreadingDynamicsBibFIXED}

\bigskip

\large{\textbf{\noindent Supporting Information}}\\
\normalsize

\bigskip

We model ring polymers using a standard bead-spring semi-flexible model based on the Kremer Grest ~\cite{Kremer1990} model. Every bead in our simulation interacts via a shifted 
Lennard-Jones potential with a cut-off $r_c = 2^{1/6} \sigma$. The gel is itself made of beads which partially overlap in order to preserve the topological status of the ring polymers, \textit{i.e.} unlinked from the the gel. The beads in the gel interact only with the beads forming the polymers via the same shifted Lennard-Jones potential. The beads forming the gel are not treated in the dynamics, meaning that the background structure is fixed and static at all times. Nearest neighbour beads along the ring polymers interact via a finitely extensible non-linear elastic (FENE) potential. The non-linear chain's flexibility is then introduced by an angular potential. 
The total intra-chain potential is therefore given by the following Hamiltonian:
\begin{align}
&H_{intra} = \sum_{i=1}^{M} \left[ U_{FENE}(i, i+1) +  \phantom{\sum_{i}^M}\right.  \notag\\
&\left. \phantom{\sum_N^M} + U_{b}(i,i+1,i+2) \right] + \sum_{i=1}^{M-1}\sum_{j=i+1}^{M} U_{LJ}(i,j) \label{eq:Hintra} 
\end{align}
where $M$ is the number of beads in each ring and the terms with $i>M$ represent those interactions needed to join the ends of the polymer in a ring fashion, \textit{i.e.} a modulo-M indexing is implicitly assumed to take into account the ring periodicity. 
Each monomer has nominal size $\sigma$ and position $\bm{r}_i$, while the distance between two monomers $i$ and $j$ is given by $d_{i,j} = | \bm{r}_i - \bm{r}_{j}|$. The finitely extensible non-linear elastic potential is of the form:
\begin{equation}
U_{FENE}(i,i+1) = -\dfrac{k}{2} R_0^2 \ln \left[ 1 - \left( \dfrac{d_{i,i+1}}{R_0}\right)^2\right]  \notag
\end{equation}
for  $d_{i,i+1} < R_0$ and $U_{FENE}(i,i+1) = \infty$, otherwise; $R_0 = 1.5$ $\sigma$, $k=30$ $\epsilon/\sigma^2$ and the thermal energy $k_BT$ is set to $\epsilon$.
The bending energy, or stiffness term, takes the standard Kratky-Porod form (discretized worm-like chain):
\begin{equation}
U_b(i,i+1,i+2) = \dfrac{k_BT \xi_p}{\sigma}\left[ 1 - \dfrac{\bm{d}_{i,i+1} \cdot \bm{d}_{i+1,i+2}}{d_{i,i+1}d_{i+1,i+2}} \right] \notag
\end{equation}
where $\xi_p$ is the persistence length of the chain which is fixed at $5$ $\sigma$. Polymers are significantly bent by thermal fluctuations at contour lengths larger than the Kuhn length $l_k = 2 \xi_p$. Here, the persistence length $\xi_p$ is always assumed to be much smaller than the total length of the chain, so that the chains resemble a flexible polymer, rather than a rigid rod. The `cut and shifted' Lennard-Jones potential takes the following form:
\begin{equation}
U_{LJ}(i,j) = 4 \epsilon \left[ \left(\dfrac{\sigma}{d_{i,j}}\right)^{12} - \left(\dfrac{\sigma}{d_{i,j}}\right)^6 + 1/4\right] \notag
\end{equation} 
for $d_{i,j} < \sigma 2^{1/6}$ and $U_{LJ}(i,j) = 0$, otherwise. The same potential is also used to regulate all the pair interactions between monomers belonging to different chains or with the fixed mesh. The inter-chain Hamiltonian takes the form:
\begin{equation}
H_{inter} = \sum_{I=1}^{N-1} \sum_{J=I+1}^{N} \sum_{i=1}^{M-1} \sum_{j=i+1}^{M} U_{LJ}(i_I,j_J)
\end{equation}
and the chain-mesh Hamiltonian:
\begin{equation}
H_{mesh} = \sum_{k=1}^{M_{gel}}  \sum_{I=1}^{N} \sum_{i=1}^{M} U_{LJ}(k,i_I)
\end{equation}
where $N$ is the number of chains in the system. The indexes $i$ and $j$ run over the beads in the chains, respectively, $I$ and $J$, and $k$ runs over the beads forming the mesh. The bead mass is $m$ and the friction acting on each bead is set to $\xi / m = \tau_{LJ}^{-1}$. The integration is performed in the over-damped limit of the Langevin equation using Verlet algorithm with time step $\Delta t = 0.01 \tau_{LJ}$, as in previous works ~\cite{Halverson2011,Halverson2011a} .

\begin{table}
\centering
\begin{tabular}{ccc|cc}
N & M & L$[\sigma]$ & $\langle R^2_g \rangle [\sigma^2]$ & $D_{CM} [\sigma^2/\tau_{LJ}]$ \\
\hline
40 & 64 & 30 & 30 & 373.9 $\times 10^{-5}$\\
20 & 128 & 30 & 60.25 & 120$\times 10^{-5}$\\
10 & 256 & 30 &	109.5 & 33.3$\times 10^{-5}$\\
\hline
50 & 256 & 50 &108.4 & 34.5$\times 10^{-5}$\\	
40 & 512 & 60 &185 & 7.67$\times 10^{-5}$	\\
50 & 1024 & 80 &296 & 2$\times 10^{-5}$	\\
50 & 1512 & 90 &390 & 8$\times 10^{-6} \downarrow$	\\
\end{tabular}
\caption{Number of chains, number of monomers, size of the system, radius of gyration $\langle R^2_g\rangle$ and diffusion coefficient of the center of mass $D_{CM}$ for the systems. The monomer density is fixed at $\rho=0.1 \sigma^{-3}$. The arrow next to the value for the longest ring system means that the value computed represents an upper bound on the real value.}
\label{tab:syspar}
\end{table}

\begin{figure}[h]
\includegraphics[scale=0.28]{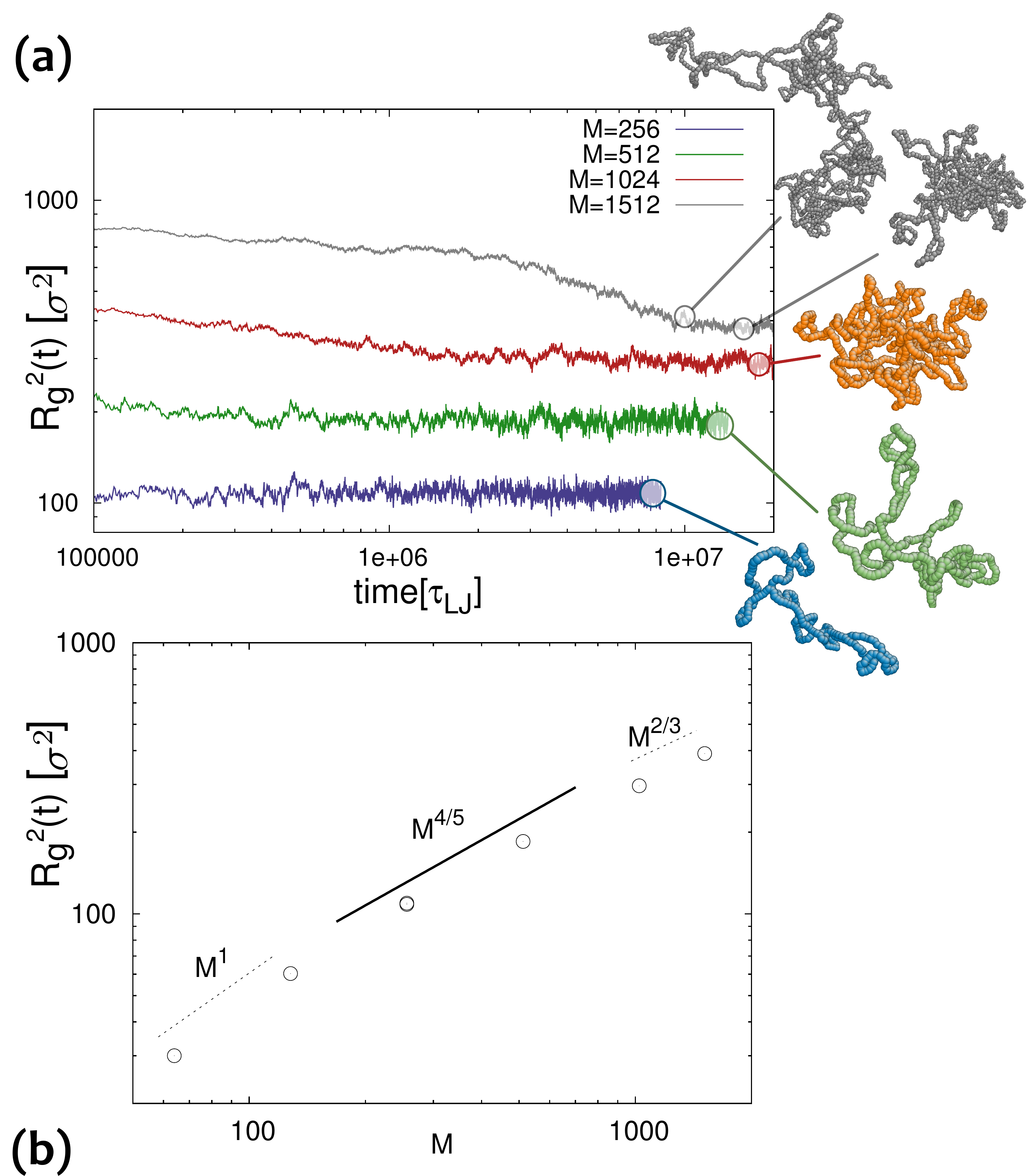}
\caption{(color online). (a) The equilibration of the gyration radius $R^2_g$ is plotted against time. Snapshots of equilibrated configurations are showed on the side. Even though our results suggest a crossover from $R^2_g \sim M$ for short rings to $R^2_g \sim M^{2/3}$ for long ones, we still observe fluctuations in $R^2_g$ which correspond to long protrusions and voids in the polymer configurations, in agreement with previous works ~\cite{Halverson2011} . (b) Scaling regimes of the radius of gyration squared. The findings reflect previous observations and theoretical predictions ~\cite{Cates1986,Halverson2011} .}
\label{fig:Rgequil}
\end{figure}

\begin{figure}
\includegraphics[scale=0.31]{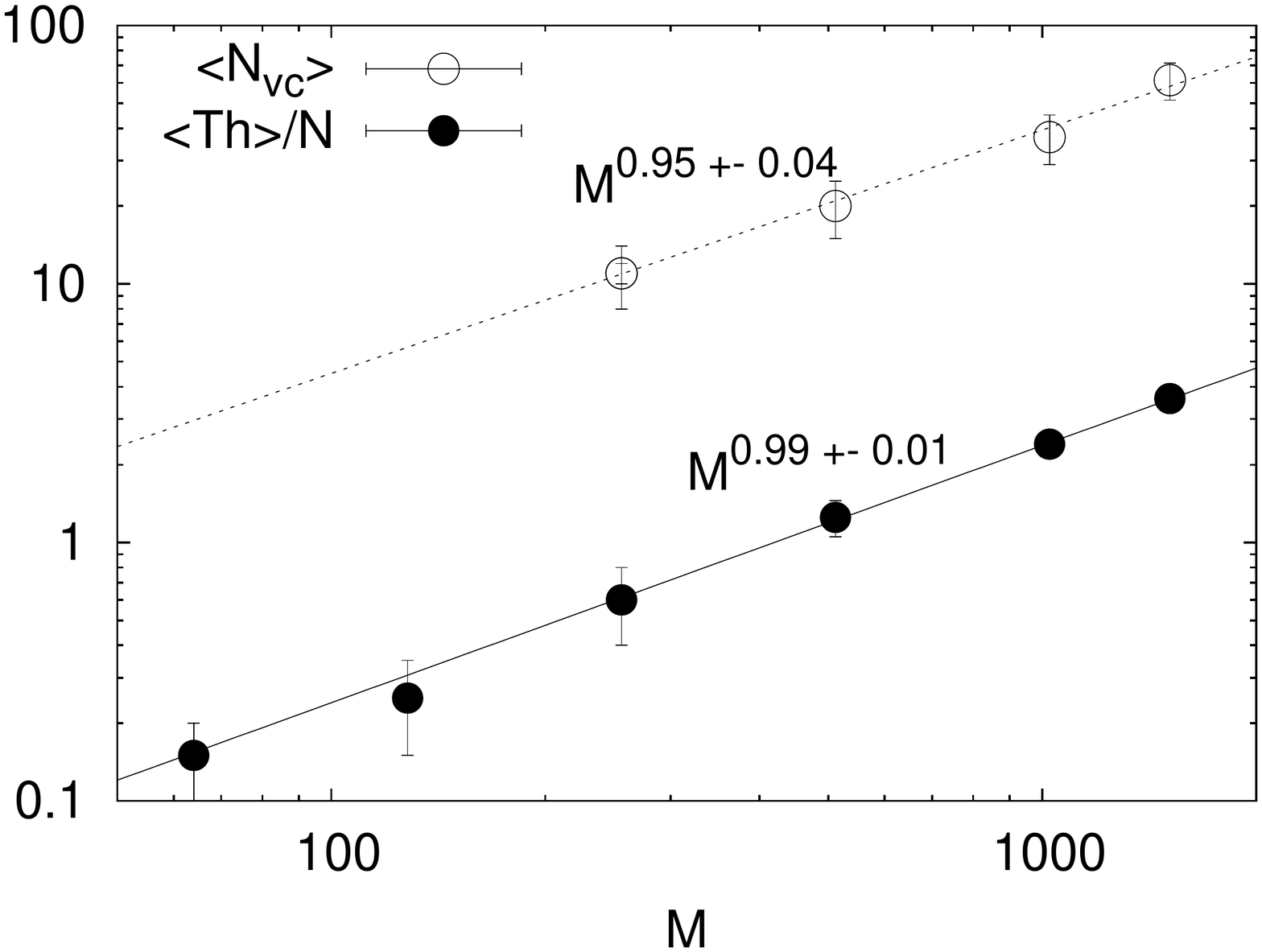}
\vspace*{-1 cm}
\caption{Scaling behaviour of the number of visited cells $N_{vc}$ by a chain  and number of threadings per chain $\langle Th \rangle/N$ as a function of the chain length $M$. Both of them show a clear linear scaling. }
\label{fig:Nvis}
\end{figure}

\begin{figure}
\includegraphics[scale=0.31]{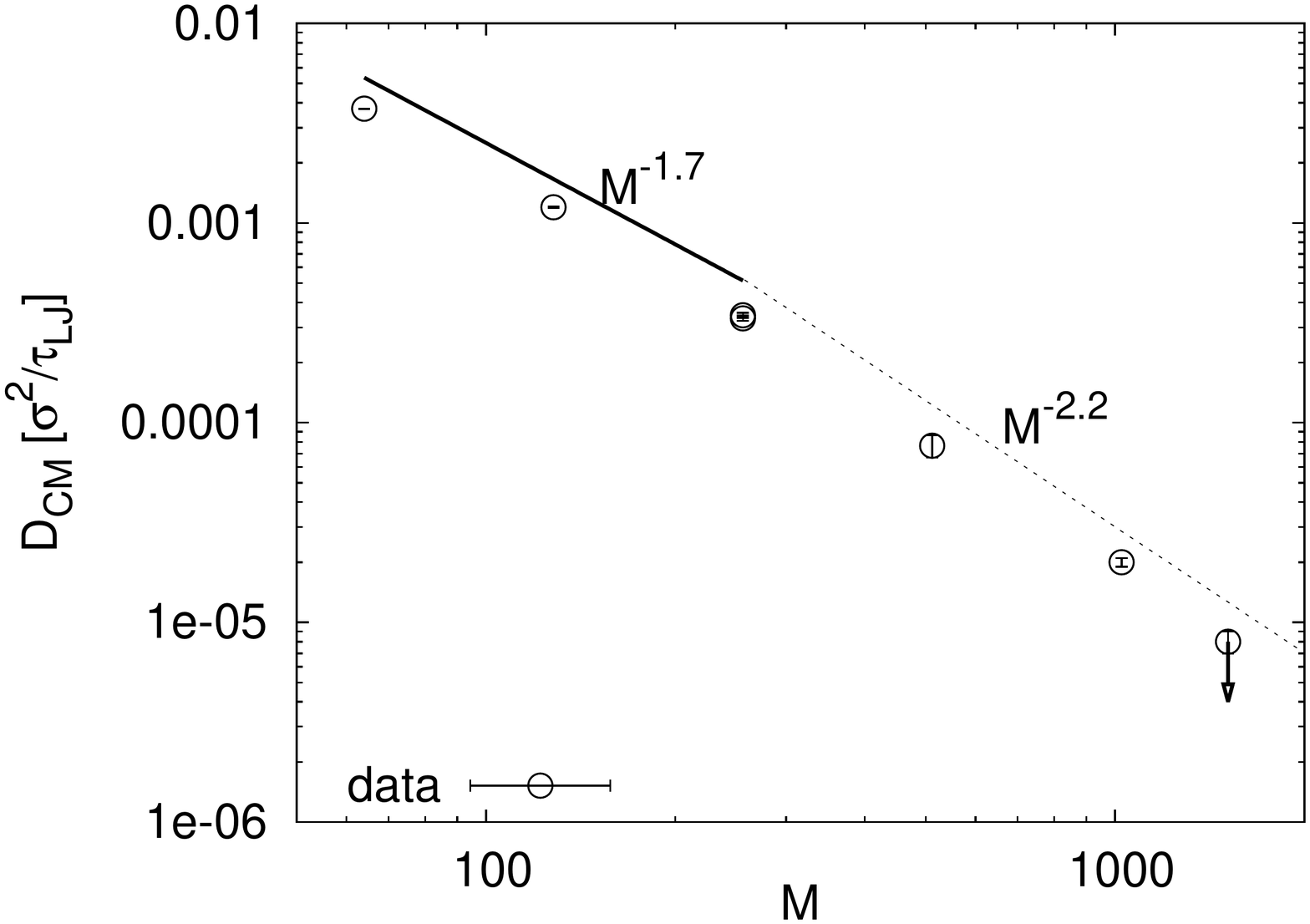}
\vspace*{-1 cm}
\caption{Diffusion coefficient of the center of mass $D_{CM}$ as a function of the chains length $M$. Error bars are smaller than the symbol size.  The arrow below the data point for the longest ring system indicates that the value computed represents an upper bound on the real value, as free diffusion is not reached within the simulation time.}
\label{fig:Dcm}
\end{figure}

\begin{figure}
\hspace*{-1 cm}
\includegraphics[scale=0.34]{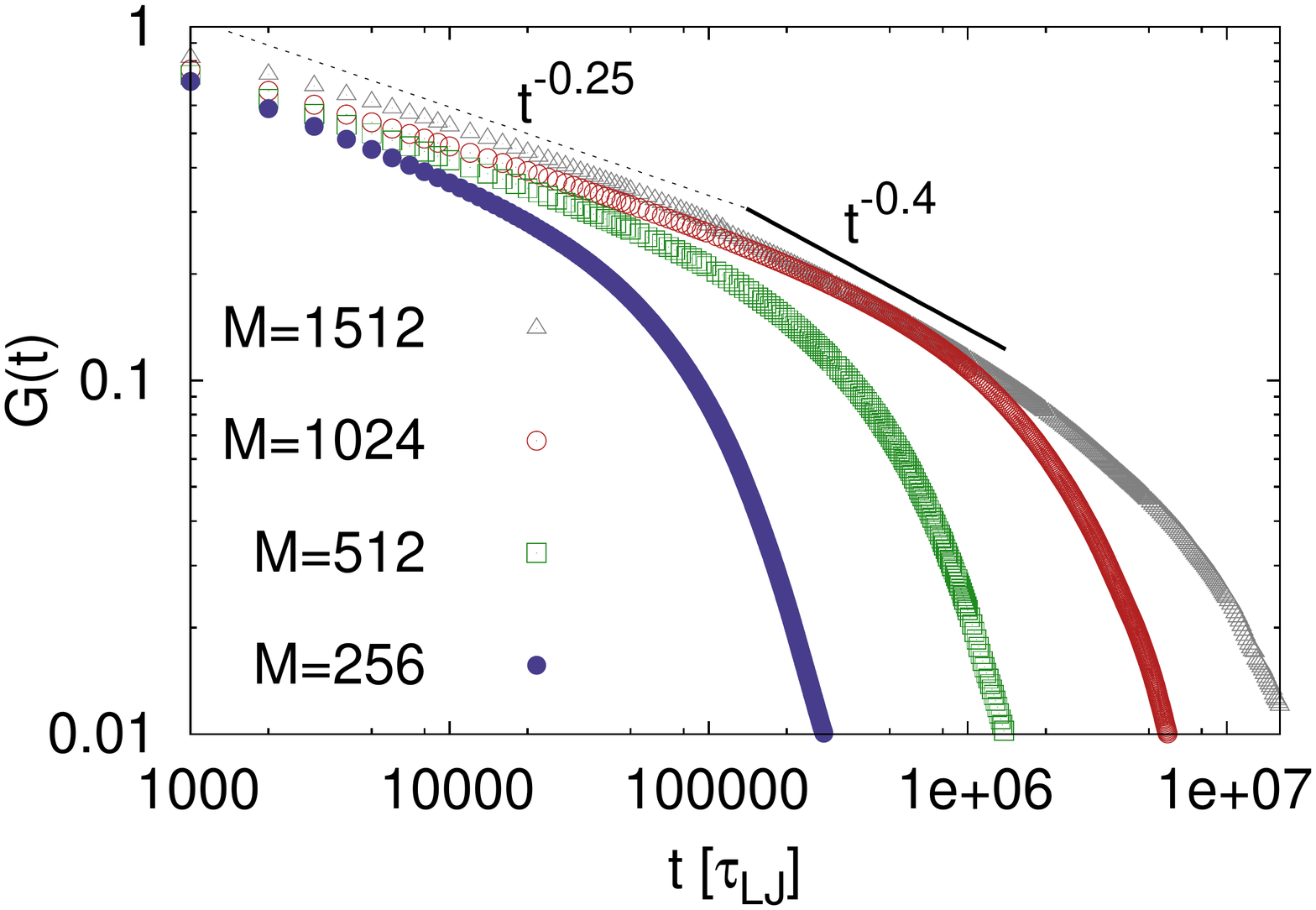}
\vspace*{-1.2 cm}
\caption{(color online). Stress relaxation modulus for systems at constant density and different chains length $M$. Rings relax the (spatial) stress much faster than their linear counterparts. The glassy plateau is also absent, in agreement with previous works ~\cite{Kapnistos2008} and ~\cite{Halverson2011a} .}
\label{fig:StrssRlx}
\end{figure}

The systems observables are averaged over the last $10^6$ timesteps for $M<1024$ and $10^7$ for $M=1024$ while for the longest rings we run a second simulation after the radius of gyration reaches a stationary state (around $10^7 \tau_{LJ}$, see Fig.~\ref{fig:Rgequil}). 
The systems are prepared in a state where the rings are folded onto themselves in order to fit through a gel channels thereby avoiding any unwanted linking, either with the mesh or other rings. Initially, the Lennard-Jones potential is turned off, and a soft-potential is used to gently push apart neighbouring non-bonded monomers. After a short run, the Lennard-Jones potential is turned on and the systems are let equilibrate until the mean squared displacement of the centre of mass of the rings travelled at least once their gyration radius. During the following simulations, we also monitored possible abrupt changes or non-stationary behaviour in the average radius of gyration of the rings in order to detect any hint of poor equilibration. We did not observe any. This is important since the rings were prepared in a very far-from-equilibrium configuration. We also checked that the rings were still unknotted after the short run performed with a soft potential pushing the monomers apart. Once the Lennard-Jones potential is turned-on, it is virtually impossible for the polymers to change their topology. The gel structure is formed by beads of size $\sigma$, which are partially overlapping in order to suppress even the point-like gaps that would appear between monomers that just touch.   
In Fig.~\ref{fig:Nvis} we report the scaling of the average number of cells visited by each chain and the average number of threaings per chain. 
We observe that both the number of visited cells $N_{vc}$ and the number of threadings per chain $Th/N$, scale extensively with the ring size $M$, as explained in detail in the main paper. 
The scaling $D_{CM}\sim M^{-\alpha}$ with $\alpha = 2$ is predicted for ring polymers ~\cite{Cates1986, Obukhov1994} . On the other hand, a crossover from $\alpha<2$ to $\alpha>2$ has been found for the melt in previous works ~\cite{Halverson2011} . We here report a similar behaviour of the diffusion coefficient of the center of mass $D_{CM}$, as shown in Fig.~\ref{fig:Dcm}. The arrow beneath the value for the longest ring systems signifies that the data point represents an upper bound on the true values since for that system, free diffusion could not be observed within the simulation time.
We also report the behaviour of the stress-relaxation modulus $G(t)$, as defined and discussed in the main paper (see Fig.~\ref{fig:StrssRlx}). Our results suggest a power law decay up to intermediate times, followed by an exponential relaxation. The value of the exponent is in agreement with the one found in ~\cite{Halverson2011a} and ~\cite{Kapnistos2008} . Two comments are appropriate. Firstly, we compute only the stress carried by the rings. The stress carried in the background gel may be significant and may be difficult to separate experimentally. Secondly, although stress relaxation in our ensemble can be compared with stress relaxation in the melt this should be done with care as there is no reason to expect the ensembles to be equivalent. Nonetheless experiments in the melt suggest an exponent close to $G(t)\sim t^{-0.5}$ rather different from the scaling we report in Fig.~\ref{fig:StrssRlx}.

\begin{figure}
\hspace*{-1 cm}
\includegraphics[scale=0.35]{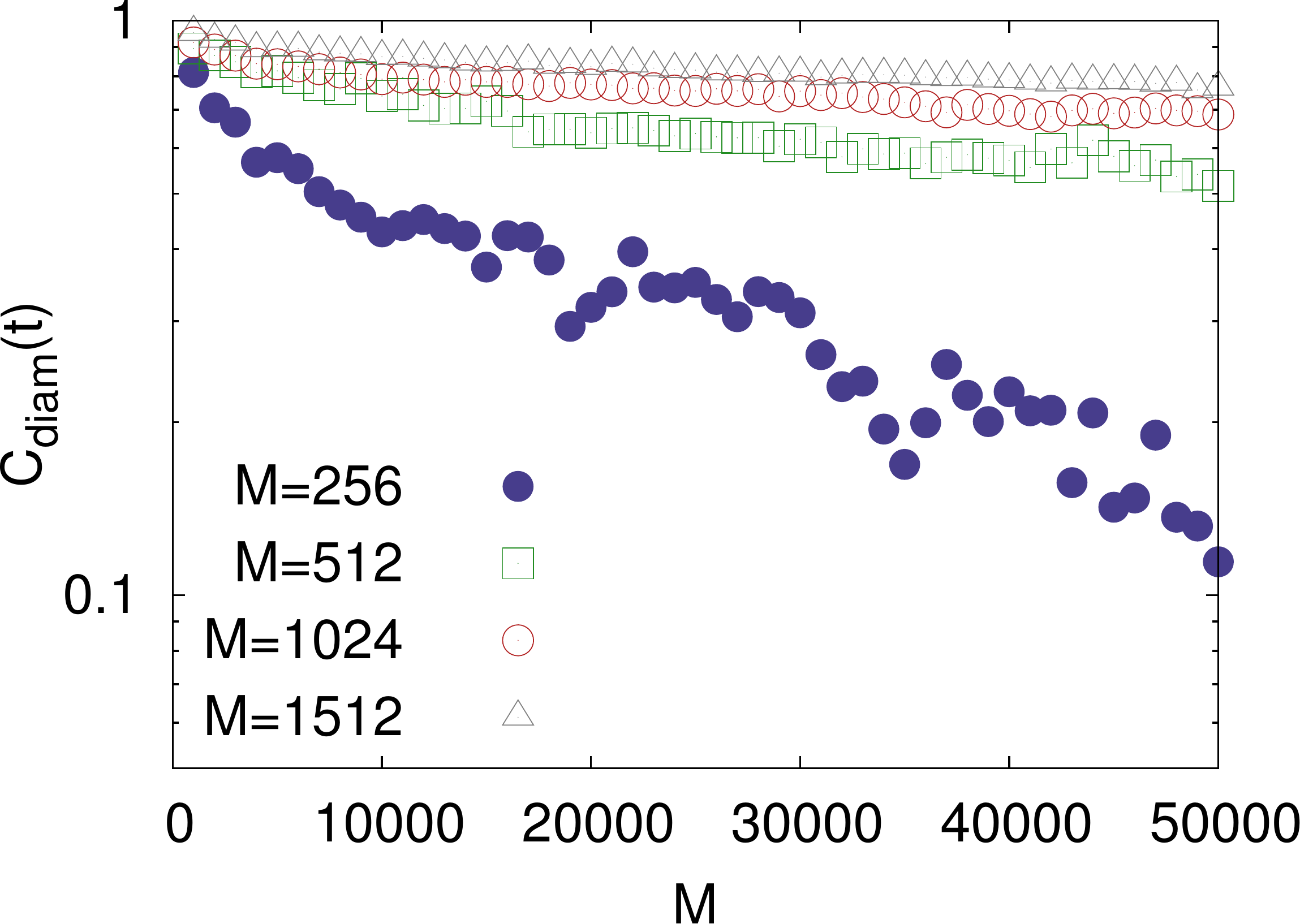}
\caption{(color online). Autocorrelation function of the polymers diameter vector $C_{diam}(t)$ as defined in eq.~\eqref{eq:Cdiam}. The behaviour suggests an exponential decay with characteristic time $\tau_{diam} = \int_0^{\infty} C_{diam}(t) dt$. } 
\label{fig:Diam}
\end{figure}

In Fig.~\ref{fig:Diam} we plot the auto-correlation function of the ring diameter vector, which is defined as: 
\begin{equation}
C_{diam}(t) = \dfrac{\langle \vec{d}(t) \cdot \vec{d}(0) \rangle}{\langle|\vec{d}(0)|^2\rangle} \label{eq:Cdiam}
\end{equation}
where $\vec{d}(t)$ is the vector joining monomers that are diametrically opposite in the chemical sense, \textit{i.e.} monomers with the largest possible chemical distance $M/2$. The average $\langle \dots \rangle$ is taken over monomer pairs and chains. The characteristic decay time, or ``re-orentation time'' (see main text), is given by $\tau_{diam}(M)$, computed as the (numeric) integral of $C_{diam}(t)$ ~\cite{Rosa2011} . 
Such a quantity gives a measure of how quickly ring polymers explore new configurations and hence how fast they relax stress. For linear polymers such a quantity is comparable to the time-scale at which free diffusion is reached. However, for ring polymers, we show in the main text that the relaxation of both penetrations (threadings) and dynamics are significantly slower than $\tau_{diam}(M)$.

As discussed in the main text one can treat $Th(i,j;t)$ as a time-dependent asymmetric adjacency matrix, associated with a directed network of penetrations. We here recall that the Betti number of a graph $G$ is defined as $b_1(G) = |\mathcal{E}| - |\mathcal{V}| + |N_{cc}|$, where $|\mathcal{E}|$ and $|\mathcal{V}|$ are the sizes of the sets of edges $\mathcal{E}$ and vertices $\mathcal{V}$, respectively, and $N_{cc}$ is the number of connected components of $G({\cal E},{\cal V})$ ~\cite{Jacques2004} . The first Betti number can be interpreted as the number of \emph{undirected} cycles of inter-threading rings or, equivalently, the number of links that can be broken without creating more connected components; it is large for a highly-connected graph and small for a tree-like structure. The Betti numbers have previously been used  to relate the jamming transition of granular media with the topology of the configurational spaces~\cite{Carlsson2011,Kondic2012} . In the main text, we used the first Betti number to relate the topology of the network of inter-threading rings to the dynamics of the polymers. The functional form of $b_1(G)$ together with snapshots of $G({\cal E},{\cal V})$ for different values of $M$ are shown in the main text. The Betti number shows a sharp transition between $M=512$ and $M=1024$ which can be interpreted as the emergence of extended inter-threading structures (see main text for details) while the strongly connected components of the graphs grow in size between $M=1024$ and $M=1512$.

\begin{figure*}
\includegraphics[scale=0.3]{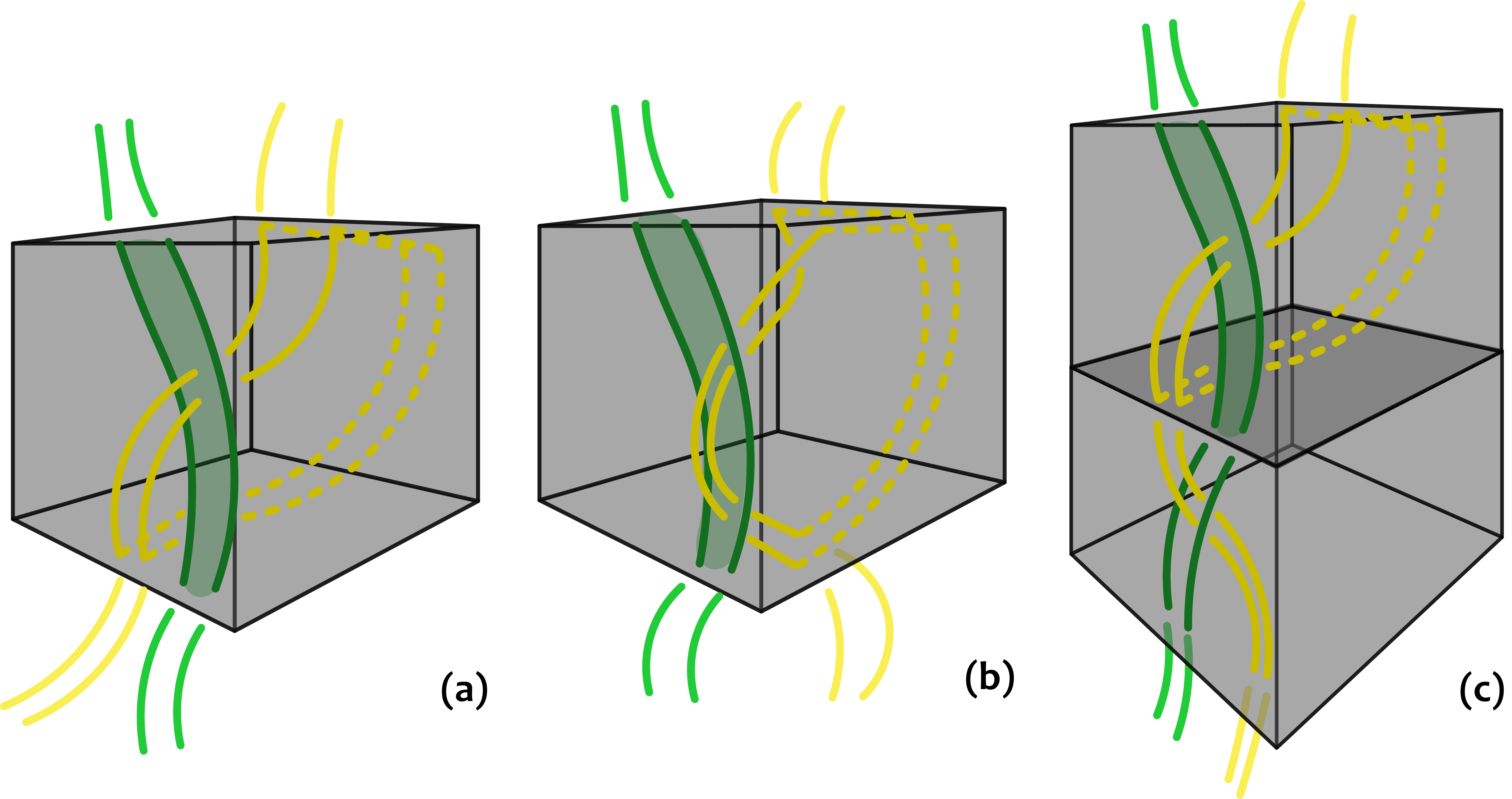}
\caption{A minority of all threadings are associated with rings that share either the entry and/or the exit faces of the unit cell of the gel, as shown here. While our definition of threading is unambiguous it is helpful to examine what is recorded in the three cases shown: (a) The rings thread through each other in the cell. Both of the yellow/yellow-dashed contours, closed at infinity are linked with the green ring, closed by a straight line connecting the points it leaves the cell. The yellow ring is therefore reported as actively threading the green ring by our algorithm. (b) There is no threading in the unit cell since the penetrating ring enters, and then leaves, the green shaded area delimited by the green ring contours. The closed yellow/yellow-dashed contours are both unlinked from  the green ring, closed on the edge of the box. 
%WHAT ABOUT MORE SPECIAL CASES?
In this case our algorithm reports three unlinked rings, hence no threading is detected. (c) As in panel (a) but one can now see that the threading is removed in the cell below in a similar manner to the situation in panel (b). In this case our algorithm detects both a local threading in the upper cell and a local threading in the bottom cell. Pairs of threadings like this will usually be short lived, as they can quickly annihilate. They will therefore not be responsible for the long lived penetrations seen in Fig. 3(a) of the main text.} 
\label{fig:ThCases}
\end{figure*}

In the main text we  explain how our algorithm defines local threadings. Here, we give a more complete explanation, describing in detail some special cases, see e.g. Fig.~\ref{fig:ThCases}. Our procedure can be divided in two steps: (1) identification of intersection points of the polymers with the faces of the unit cell, (2) construction of closed contours based uniquely on the information contained \emph{inside} the  unit cell. We choose to adopt this local procedure since the global topology, being unlinked, would not give any insight regarding penetrations, reflecting the difficulty of identifying them in ring polymer melts (or solutions). Our definition allows us to quantify the local constraints that the rings exert on each-other on the scale of a lattice volume in the the form of penetrations (threadings). Our algorithm first generates one loop from the in-coming and out-going segments of the same polymer ring (green in Fig.~\ref{fig:ThCases}) , closed by straight lines across the faces. Next, two closed contours are generated from each segments of all other polymers in the cell (one of which is shown yellow in Fig.~\ref{fig:ThCases}). If the last two are linked with the first we identify this as a local threading, by definition. Such a procedure relies only on the configuration/information inside the unit cell, and therefore returns local information only. In Fig.~\ref{fig:ThCases} we show three cases as an example. In Fig.~\ref{fig:ThCases}(a) the yellow ring penetrates the green one inside the cell, (b) the yellow ring does not thread completely through the green one and (c) the yellow ring threads locally in one direction in the upper cell and then in the opposite direction in the lower cell. Our algorithm detects both local threadings.

\newpage

\bibliography{ThreadingDynamicsBibFIXED}

\end{document}